\documentclass[a4paper,british]{scrartcl}

\usepackage[T1]{fontenc}
\usepackage[latin9]{inputenc}
\usepackage{graphicx}

\makeatletter

\providecommand{\tabularnewline}{\\}

\makeatother

\usepackage{babel}
\usepackage{listings}

\begin{document}
\title{Finding Experts in Social Media Data using a Hybrid Approach}
\author{Simon James (Seamus) Brady}
\maketitle
\begin{center}
\par\end{center}

\thispagestyle{empty}
\clearpage 
\pagestyle{plain}

\pagebreak{}

\tableofcontents{}

\pagebreak{}

\listoffigures

\listoftables

\pagebreak{}

\section*{Abstract}

Several approaches to the problem of expert finding have emerged in
computer science research. In this work, three of these approaches
- content analysis, social graph analysis and the use of Semantic
Web technologies are examined. An integrated set of system requirements
is then developed that uses all three approaches in one hybrid approach.

To show the practicality of this hybrid approach, a usable prototype
expert finding system called ExpertQuest is developed using a modern
functional programming language (Clojure) to query social media data
and Linked Data. This system is evaluated and discussed. Finally,
a discussion and conclusions are presented which describe the benefits
and shortcomings of the hybrid approach and the technologies used
in this work.

\pagebreak{}

\clearpage 

\section{Introduction}

The original idea for this report arose from my frustration at endeavouring
to find particular expertise in search engines and a growing realisation
that expert finding was actually a much discussed and researched computer
science topic. After some research, the initial aim was to build an
expert finding system using Semantic Web technologies but the research
suggested that this topic had already been covered in some depth.
What was missing from the research was a practical examination of
what sort of expert finding system could be developed with the data
and tools as existing on the Internet today.

This report is an attempt to build this system, a practical, usable
prototype expert finding system called ExpertQuest. This system will
collect data from Twitter, GitHub and DBPedia to help users find computer
programming experts.

\subsection{Objective}

The objective of this report is to see if it is possible to build
and test a usable prototype expert finding system within the following
constraints:
\begin{itemize}
\item Using a modern programming language.
\item Using only data that is available on the Internet via social network
APIs and Linked Data.
\item Uses a mixed or hybrid approach to expert finding.
\item Fulfil the requirements as outlined in Chapter 3.
\end{itemize}

\subsection{Structure of the Document}
\begin{itemize}
\item Chapter 2 discusses the need and value of expertise, as well as how
expertise is defined and how expertise can be managed via technology.
\item Chapter 3 outlines the particular strategy and requirements used in
the system implemented for this report.
\item Chapter 4 discusses various technology and design choices for implementing
the expert finding system.
\item Chapter 5 contains a detailed description of the actual design and
implementation of the expert finding system.
\item Chapter 6 outlines the testing and validation carried out on the results
of the expert finding system.
\item Chapter 7 outlines conclusions and some ideas for future work.
\end{itemize}
\pagebreak{}

\section{Background}

This chapter will discuss the need for and value of expertise, as
well as how expertise is defined and how expertise can be managed
via technology.

\subsection{The Need for Experts}

\subsubsection{Experts in the Wild}

In 1960 the famous primatologist Dr. Jane Goodall observed that chimpanzees
used specially chosen twigs to fish for termites \cite{1}. This was
the first time a scientist had observed this behaviour and it helped
open new avenues of research into animal tool use. 

Since then, tool use has been observed in seven classes of the animal
kingdom, including primates and crows \cite{2}. One of the most interesting
things about the research into chimpanzee tool use is that some of
the tools that chimps use are ``culture-bound''. Only certain tools
are known to a particular populations of chimps:
\begin{quote}
``Many differences can be found in tool behaviour for which neither
genetic nor ecological explanations are suitable.'' \cite{3}
\end{quote}
The use of tools is quite often taught by one generation to the next
generation, so that in a sense these chimps have a local tool culture
\cite{4}. This also implies that certain individual chimps in a troop
have more knowledge about certain tools than some other chimps. The
use of tools allows these chimps to broaden their ability to adapt
and thrive by opening up new sources of nutrition. These local chimp
``experts'' are helping to keep the troop alive. 

This is much the same as our species case - human beings are also
reliant on the skills and knowledge of multiple individuals for our
survival. 

\subsubsection{The Expert Finding Problem}

In a large organisation, perhaps even global, finding someone who
is an expert may be a challenge - especially when compared to finding
the local expert in a small neolithic community.

The problem can be called the \textit{expert finding problem} or \textit{expertise
location problem}. We shall stick with the term \textit{expert finding}. 

As may be expected, there is a large body of software and research
available on the expert finding problem and even the researchers do
not agree on an exact definition.

This problem is defined by Lappas (2011) quite simply:
\begin{quote}
``... given a task at hand and a set of candidates, one wishes to
to efficiently identify the right expert (or set of experts) that
can perform the given task.'' \cite{5}
\end{quote}
In their research, Yimam-Seid and Kobsa (2003), also found that expertise
is sought out for two different reasons. They identified:
\begin{quote}
``... two main motives for seeking an expert, namely as \textit{a
source of information} and as \textit{someone who can perform a given
organizational or social function.'' \cite{6}}
\end{quote}
These two definitions can be combined into the following single definition
of the expert finding problem which will be used in this thesis:\textbf{
}\textit{Efficiently identify the right individual (or group) from
a field of candidates that has the expertise to provide desired information
or complete a desired task.}

This definition gives a viewpoint from which to survey the various
strategies that have been used to solve the expert finding problem
using software. For the sake of this thesis, it will be assumed that
returning an individual (rather than a group) is sufficient.

\subsubsection{The Evolution of Human Expertise}

The story of human evolution is also wrapped up in the evolution of
human expertise. Indeed, there is some evidence to suggest that our
ability to think and communicate socially as human beings is intimately
wrapped up in the evolution and increasing complexity of our tool
use:
\begin{quote}
``Lower Palaeolithic technologies clearly do increase in hierarchical
complexity through time, raising the possibility of important interactions
with the evolution of human cognitive control and socially supported
skill acquisition.'' \cite{7}
\end{quote}
Human expertise has had a profound impact on our own species, even
to the point that increasing human expertise in tooling has improved
our cognitive and social abilities. 

\subsubsection{Expertise and the Global Corporation}

As the global economy has become more complicated, expertise has become
more and more valuable in the modern corporation. The tangible results
of having experts working within your company include patents, copyrighted
works, trade marks and other forms of what are known as \textit{intellectual
property}. In fact, the true value of intellectual property within
a business may be worth more than all of the physical assets combined:
\begin{quote}
``During 2000, the market-to-book ratios of Fortune 500 companies
increased to 6.3:1, suggesting that for every dollar of physical assets
on the balance sheet, the market recognized \$6.30 worth of other
intangible assets. 

On average, in successful organizations, brands, intellectual property
and the like are two to three times the value of physical assets.
Intellectual property holdings are valuable corporate assets, and
may make up a great portion of the total worth of an organization.''
\cite{8}
\end{quote}
With this new level of value attached to expertise, and the importance
of experts to business continuity (such as after a major disaster),
there now exists an even greater desire to identify, manage and co-ordinate
the expertise within an organisation. 

\subsubsection{What is Expertise?}

So experts and expertise have had a profound impact on human beings
and their closest primate relatives and experts are also important
in the economy. But what exactly does the word ``expert'' mean?

The word ``expert'' generally means someone who knows much about
a specific subject matter or discipline. Expertise is the noun describing
this quality. However, the concept is more subtle than that. 

The online Business Dictionary defines ``expertise'' in the following
way:
\begin{quote}
``Basis of credibility of a person who is perceived to be knowledgeable
in an area or topic due to his or her study, training, or experience
in the subject matter.'' \cite{9}
\end{quote}
This definition offers a slightly more nuanced view of expertise:
\begin{itemize}
\item Expertise is a quality of an expert, a person that has studied, trained
and gained experience in a particular subject matter. This person
has more skills and knowledge than most others in a specific area.
\item This person is perceived to have this quality by other people - expertise
has a social dimension. A hermit can know everything about growing
orchids, but without a society to communicate with, nobody will be
able to consider the hermit as an expert (apart from the hermit of
course, but this is perhaps tautological).
\item It is also implied that this person has credibility or is trusted
by others who perceive this expertise. Again, if our hermit suddenly
appears in the village square claiming to be an expert orchid grower
but refuses to supply any kind of bona fides, the villagers would
probably be correct to assume that he or she is a crank.
\end{itemize}
So we can say that expertise has a knowledge dimension, a social dimension
and a trust dimension. To be fully considered an expert, a candidate
must have the necessary knowledge, have a group of peers aware of
this knowledge and have a level of credibility amongst these peers
that this knowledge is valid. These properties are summarised in \textbf{Table
1}. 

These properties will provide a useful framework for the discussion
of expert finding. We shall use these three properties to judge several
approaches to expert finding in the following sections. Each of the
expert finding approaches we are using can be compared using the following
questions:
\begin{itemize}
\item (1) How does this approach measure ensure that the expert has adequate
knowledge?
\item (2) Does this approach measure the social profile of the expert in
a meaningful way to show that the expert is regarded as an expert
by his/her peers?
\item (3) Does this approach validate the credibility expertise of the expert
in any way?
\end{itemize}
There is some crossover between question 2 and 3 (as large groups
of peers who regard a candidate as an expert does strengthen the candidate's
credibility). However they will be treated separately as there is
enough divergence in other forms of expertise validation (education
or research for instance) for the distinction to be useful. 

To these three questions we can also add a fourth:
\begin{itemize}
\item (4) Does this approach efficiently identify the right individual from
a field of candidates that has the expertise to provide desired information
or complete a desired task?
\end{itemize}
The word ``efficiently'' does have some importance here, as speed
and ease of use should be considered when comparing approaches.

\begin{table}
\begin{centering}
\begin{tabular}{|l|l|}
\hline 
\textbf{Adequate Knowledge} & A candidate must have the necessary knowledge\tabularnewline
\hline 
\textbf{Peer Awareness / Social Profile} & A candidate has a group of peers aware of their knowledge\tabularnewline
\hline 
\textbf{Peer Credibility} & A candidate is trusted by their peers\tabularnewline
\hline 
\end{tabular}
\par\end{centering}
\caption{\textbf{The Three Properties of Expertise}}
\end{table}

\subsection{Expert Finding Using Content Analysis}

\subsubsection{Introduction}

\textit{Content analysis} is the approach used by much of the expert
finding academic work found in the Text REtrieval Conferences (TREC)\cite{10}.
In this work, content analysis is taken to be a synonym of textual
analysis, and the phrases are used interchangeably.

This work was the mainstay of expert finding research until the rise
of the social Web (which will be examined in the next section). In
his survey of expert finding techniques, Lappas et al (2011) call
this approach ``Expert Location Without Graph Constraints''. 

The content analysis usually amounted to using textual analysis techniques
on large data corpora, for instance the email archives of any large
organisation. In fact, the email list archives of the World Wide Web
Consortium was often used as a basis for academic research as they
were publicly available. 

This approach concentrates on estimating
\begin{quote}
``the probability that a candidate... could be an expert with respect
to a given topic query''\cite{11}
\end{quote}
In order for a data corpus to be analysed, it would need to be put
through standard text preprocessing:
\begin{itemize}
\item Sanitising the data.
\item Tokenising the remaining data.
\item Stop word removal and stemming the words in the corpus.
\end{itemize}
After this, there are two possible models that can be pursued (Balog
et al, 2006 \cite{12}):
\begin{itemize}
\item A \textit{candidate-centric} model where the various documents associated
with a candidate are grouped and then a vocabulary of terms is extracted
from the documents to represent the candidate's expertise. The probability
of the set of terms in the query being represented in each candidate's
model can be then generated using Bayesian analysis. 
\item A \textit{document-centric} model is where the set of terms for each
document are extracted and ranked based on the terms in the query.
The candidates associated with each returned document are then found
and an aggregated set of terms for all documents belonging to this
candidate are generated. This set of terms is then analysed against
the query to return the top candidate using Bayesian analysis. 
\end{itemize}
Balog found that the document-centric model outperformed the candidate-centric
model. One of the main reasons is that an index of candidates did
not need to be kept up to date as the list of possible candidates
could be generated dynamically from the list of documents returned
from the initial query.

There are other variations on this approach but they mainly retain
the same emphasis on modelling expertise based solely on the set of
terms available in the text of the data corpus. One such variation
was produced by Serdyukov and Hiemstra (2008) where terms associated
with a candidate were augmented with sets of terms found in various
online searches:
\begin{quote}
``We used various kinds of GlobalWeb search services to acquire a
proof of expertness for each person which was initially pre-selected
by an expert finding algorithm using only organizational data.''
\cite{13}
\end{quote}
While this is an interesting avenue of research, it has the same emphasis
on text based analysis.

\subsubsection{Content Analysis and Adequate Knowledge}

Content analysis measures the expertise of each candidate by modelling
the expertise of candidates using standard, well understood information
retrieval techniques. Various different types of content for example,
emails, academic papers, business documents, can be combined into
one system for analysis. Other types of text-based information such
as Web based searches can be used to augment the system.

\subsubsection{Content Analysis and Social Profile}

The main disadvantage of content based expert finding is that it only
addresses the first of the three properties of expertise as outlined
in \textbf{Table 1}, \textit{Adequate Knowledge}. This knowledge is
captured in the set of terms associated with each candidate. As noted
earlier, Lappas (2011) called this approach ``Expert Location Without
Graph Constraints'' and for a good reason. There is very little analysis
of the social profile of the expert. There is not much knowledge of
where the expert appears in the social graph of his or her peers. 

It can be difficult or impossible to gain this type of knowledge based
solely on textual analysis of terms. Of course, this knowledge may
not be sought (as when a user searches in a search engine, the popularity
of an expert is secondary to the expertise being provided). However
there are situations where the social profile of an expert may be
useful, for instance, language or other cultural factors, when all
other criteria are equal.

\subsubsection{Content Analysis and Expertise Credibility}

For much the same reasons as above, the actual credibility of the
expertise of the candidate amongst his or her peers is difficult to
assess also, based solely on textual analysis. What if the documents
supplied have been falsified, for instance?

\subsubsection{Content Analysis and Effective Identification of Experts}

This approach can identify an expert from a field of candidates, once
the caveats above are considered. However one disadvantage is that
the data corpus may have to go through extensive extraction and transformation
before it can be used, for example as in Balog et al, 2006 \cite{14}.
Once this ETL (Extract, Transform, Load) process is automated, the
system may be efficient but there is an upfront development cost for
this data munging.

\subsubsection{Commercial Expert Finding Software Using Content Analysis}

There have been many commercial expert finding systems available for
purchase, normally by large corporations with extensive email and
document corpora that could be mined for information. Some of these
systems were reviewed by Maybury et al (2002) \cite{15}. At the time
of writing of Maybury et al's research (2002), most of these systems
would have fitted into the content analysis grouping.

\subsection{Expert Finding Using Social Graph Analysis}

\subsubsection{Introduction}

Since the rise of social media, much of the academic research on expert
finding has been concentrated on augmenting the results of textual
analysis with an \textit{analysis of the social graph} to discover
if information about expertise can be gleaned from the links between
people. Lappas et al (2011) call this approach ``Expert Location
with Score Propagation''. They describe it as a two stage process:
\begin{quote}
``1) using language model or heuristic rules {[}to{]} compute an
initial expertise score for each candidate... and 2) using graph-based
ranking algorithms to propagate scores computed in the first step
and rerank experts...'' \cite{16}
\end{quote}
This research improves on the results of textual analysis - as Bozzon
et al (2013) observed about their own research, it showed: 
\begin{quote}
``the empirical demonstration of the greater contribution of activities
of social network members with respect to their profiles for assessing
the user expertise. We also found that certain profiles and activities
of closest social contacts may provide useful information, thus giving
a positive contribution to the expert ranking.'' \cite{17}
\end{quote}
There are several approaches that can be used to do this scoring and
each normally revolves around describing the algorithm that is used. 

A variation of the PageRank algorithm was proposed by Kardan et al
(2011) \cite{18}. They called their variation the SNPageRank Algorithm.
The PageRank algorithm is famously used by Google and can be simply
described as a link counting algorithm. The more links a website has,
the greater the importance of the website. Kardan et al extended this
idea to social networks, so rather than counting links, their algorithm
counted posts, likes and connections between people in a social network
(Friendfeed). The more messages a person had, the more important they
were in the network. They extracted data from social networks and
analysed it using their modified algorithm. They found that they could
achieve positive results by using this approach.

The Hyperlink-Induced Topic Search (HITS) Algorithm was developed
by Jon Kleinberg for ask.com \cite{19}. This was examined by Dom
et al (2003) \cite{20} as a possible expert finding algorithm. This
algorithm is normally used to measure Web page ``authority''. A
Web page becomes an authority if it has many links from what are called
``hubs'', or Web pages that links to a lot of other authority Web
pages. They used the HITS algorithm on an email corpus. They considered
those who received lots of email enquiries to be ``authorities''
and those who forwarded on queries to be ``hubs''. In this particular
instance, the algorithm was not as successful as others, possibly
because email communication is not naturally structured in a strong
hub versus authority way.

\subsubsection{Social Graph Analysis and Adequate Knowledge}

An initial phase of textual analysis generally provides a measure
of the expertise of the candidate. This is used to generate profiles
for each candidate based on the results of the textual analysis.

\subsubsection{Social Graph Analysis and Social Profile}

Social graph analysis can provide insight into the social profile
of the expert, which may lead to further insights into their expertise
(for instance, a particular age profile or profession may be associated
with one expert and not another). The main impact of social graph
analysis is that, depending on the algorithm, one can measure the
number and strength of links to the expert, allowing comparison of
link scores.

\subsubsection{Social Graph Analysis and Expertise Credibility}

The main advantage of using social graph analysis is that two of the
main properties of expertise can be analysed. An extensive social
profile can be a good indicator of expertise. However, an extensive
social profile does not guarantee expertise per se. For instance,
a mediocre expert may gain followers on social media due to good public
relation skills, rather than actual expertise. Other factors may also
need to be assessed for ensuring expertise credibility, for example
peer reviewed science papers.

\subsubsection{Social Graph Analysis and Effective Identification of Experts}

The social graph can indeed provide information about experts. The
ironic thing about link analysis and expert finding is that outside
the world of computer science research, social media is already used
to find expertise everyday:
\begin{quote}
``Social networking plays a significant role in expert finding. Often
we require a more general expert to suggest the specialised expert
we need by referring to colleagues in his or her social network. Social
networking also helps to find an expert by providing a group of people
within a community and perhaps links of people outside of the community
as well.'' \cite{21}
\end{quote}
As with the textual analysis approach, there may be an upfront cost
in terms of data extraction and transformation e.g. Kardan et al (2011)
\cite{22}. Also, when compared to the textual analysis approach where
a data corpus may be available on a local network, the distributed
nature of the social graph may introduce its own complications in
terms of network latency.

The success of link or social graph analysis may also depend on the
choice of algorithm. As Dom et al (2003) found, certain algorithms
may not fit the ``shape'' of certain data corpora and may not produce
conclusive results. Also, privacy may become a major issue when using
a social network data to find expertise.

\subsection{Expert Finding Using Semantic Web Technology}

\subsubsection{Introduction}

There is also a subset of expert finding research that uses \textit{Semantic
Web technology}. This approach overlaps with the content analysis
and social graph analysis approaches but it is worthy of examination
in its own right. Notably Lappas et al (2011) \cite{23} do not mention
any of this research in their excellent survey. 

In the literature there are very few outlines of expert finding systems
that actually use Semantic Web technology. One of the few is Li at
al (2006) in which they describe a system called FindXpRT. This system
was developed to use the FOAF ontology (Friend of a Friend ) \cite{24}
with an extra series of rules implemented in RuleML \cite{25}, an
XML based rule deduction markup language. The experts' profiles in
the system were implemented in FOAF as FOAF facts. Various rules in
RuleML were then used to augment these facts - rules to allow clients
to find an expert to collaborate with or rules that will allow an
expert to pick another expert to collaborate with. 

In their survey paper, Titus Schleyer et al (2008), mention a couple
of challenges of using Semantic Web technology in an expert finding
system, specifically for finding biomedical experts. Among these problems
were the fact that Semantic Web technology does not meet all the requirements
of building an expert finding system and it would likely be:
\begin{quote}
``a useful technological infrastructure for implementing expertise
location systems, not as an end-to-end architecture'' \cite{26}.
\end{quote}
However Schleyer et al do recommend Semantic Web technology for keeping
an expert's profile up to date by linking it to their activities across
the Web. They also suggest connecting disparate users and expertise
domains in useful ways using specially adapted ontologies and using
data aggregated by other Semantic Web technologies to augment existing
data stores.

This use of Semantic Web as an augment to other approaches to expert
finding is also echoed in Stankovic et al (2010). They specifically
discuss Linked Data (LOD is defined as Linking Open Data):
\begin{quote}
``Traditional approaches tend to retrieve their data from closed
or limited data corpuses. LOD on the other hand allows querying the
whole Web like a huge database, thus surpassing the limits of closed
data sets, and closed online communities. We believe that this opens
new possibilities for traditional expert search and profiling systems
which usually only rely on data from their local and limited databases
or on unstructured data gathered from the Web. LOD also stands up
for a great promise to deliver multipurpose data that can be used
to find experts in many domains and with many different expertise
hypotheses.'' \cite{27}
\end{quote}
However, Stankovic et al also mention that there are challenges with
merging duplicate data sets, verifying the authenticity of expertise
and standardising how expertise data, for example science papers,
are captured. All of these challenges arise out of the distributed
nature of Linked Data.

\subsubsection{Semantic Web and Adequate Knowledge}

One of the main advantages that Semantic Web technologies offer is
a way out of the information silos that expertise data tends to end
up in. With the correct use of common ontologies and Linked Data,
expertise from multiple disciplines could be merged and queried. Semantic
Web technologies offer the possibility of using richer data sources
to generate candidate profiles.

\subsubsection{Semantic Web and Social Profile}

Several Semantic Web technologies can be used to measure the social
profile of a candidate, for example the FOAF protocol.

\subsubsection{Semantic Web and Expertise Credibility}

As was pointed out above, the possibility of rich connected data sources
to generate candidate profiles could allow several different checks
to be made against a candidate's bona fides, thereby increasing the
credibility of each candidate profile.

\subsubsection{Semantic Web and Effective Identification of Experts}

Semantic Web technologies do not identify experts on their own. Semantic
Web technologies show major potential for linking disparate areas
of expertise for analysis and searching. However, as Titus Schleyer
et al (2008) \cite{28} point out, there needs to be a technological
infrastructure in place to take advantage of these technologies and
as yet, these types of developments are not common. So the potential
of Semantic Web technologies in expert finding remains in the area
of academic study, unlike for instance content analysis, where several
commercial systems exist using the content analysis approach.

\subsection{Summary}
\begin{itemize}
\item Experts are a vital part of both human culture and the modern corporation. 
\item Expert finding can be defined as the problem of identifying the right
individual from a field of candidates that have the expertise to provide
desired information or complete a desired task.
\item Expert finding is a challenging but necessary problem in managing
intellectual property. 
\item Expertise can be said to have a knowledge dimension, a social dimension
and a credibility dimension.
\item The content analysis approach offers great strengths in providing
the knowledge dimension of a candidate.
\item The social analysis approach can be used to measure the social profile
of a candidate.
\item Semantic Web technologies can be used to augment the other two approaches,
including increasing the credibility of a candidate by querying multiple,
rich data sources.
\item No single approach provides a complete solution to the expert finding
problem.
\end{itemize}
\pagebreak{}

\section{Strategy and Requirements}

The rest of this document will outline the design, development and
testing of a prototype expert finding system that will be called \textit{ExpertQuest}.
This chapter outlines the particular strategy and requirements used
in the system implemented for this report.

\subsection{Definition of the Hybrid Approach}

ExpertQuest is inspired by, rather than based on, the approach by
Metze et al (2007) when they designed the Spree System. They describe
the basic system as:
\begin{quote}
``a system to facilitate exchange of information by automatically
finding experts... {[}our{]} objective is to provide an online tool,
which enables individuals within a potentially large organization
to search for experts in a certain area, which may not be represented
in company organization or reporting lines'' \cite{29}
\end{quote}
The actual Spree system \cite{30} consists of a Python Web application
that uses ontological data derived directly from the Web (Yahoo Search
API) in conjunction with an algorithm that uses this data to classify
experts. The website then uses a community of users to further classify
expertise as they interact with each other. This is what is meant
by the \textit{hybrid approach}. Spree uses all three of the approaches
that were outlined in the previous chapter in one system - content
analysis, social graph analysis and Semantic Web technology. This
gives us a design approach that can be used for the ExpertQuest system.
The system should be Web based and use all three approaches to expert
finding together, in order to fulfil the primary and secondary requirements
outlined below.

\subsection{Primary Requirements}

As part of this development, the ExpertQuest system will attempt to
fulfil the key requirements that Maybury et al \cite{31} outline
for an expert finding system:
\begin{enumerate}
\item The system should be able to identify experts from a field of candidates.
\item The system should be able to classify the level of expertise of each
candidate.
\item The system should be able to validate the expertise of each candidate.
\item The system should be able to rank candidates on multiple dimensions.
\end{enumerate}
These will be regarded as the primary requirements to evaluate the
system. 

\subsection{Secondary Requirements}

Alongside these core requirements, ExpertQuest should also meet the
following secondary requirements:
\begin{enumerate}
\item The system should use the hybrid approach as defined above and measure
in some meaningful way, all three dimensions of expertise as outlined
in the last chapter - the knowledge dimension, social dimension and
credibility dimension.
\item The system should be based on real-time data available on the Internet
as much as possible. This will help to avoid the problems associated
with stale expert profiles. Any information on an expert should be
up to date. The system should also minimise any need for expensive
extraction and transformation of data such as outlined by Kardan et
al (2011). There will be no delays for batch processing data or anything
similar, at least in the prototype.
\item The system should also allow a way to contact the expert if possible.
\item The system should be as real-world as possible, not just a theoretical
outline - basically a usable prototype.
\item If possible, the system should be user friendly and reasonably fast
and should have a Web interface.
\end{enumerate}
These secondary requirements will also be used to evaluate the system.
The design choices and constraints arising from these requirements
will be discussed in the next chapter.

\pagebreak{}

\section{Design Choices}

This chapter discusses various technology and design choices for implementing
ExpertQuest. 

\subsection{Computer Programming Expertise as Domain for Expert Finding}

Many areas of expertise were considered as the domain for ExpertQuest
and even making ExpertQuest handle multiple domains. However, the
multiple domain idea was quickly dismissed as too complicated for
this document. Moreover, several domains stood out in terms of available
data, and the following domains were isolated as possibilities:
\begin{itemize}
\item Biotechnology
\item Medical and genetic research
\item Computer Science
\end{itemize}
Each of these would be an excellent candidate as a topic, but in the
end the the author's experience in computer programming made this
the obvious choice as a domain of expertise. A knowledge of the technical
jargon in the field and a rich availability of data sources made this
an easy choice. It was also decided that confining the system to one
subset of a domain (computer programming versus all of computer science)
would make the system easier to build and evaluate in the time available.

However, there were other factors in making computer programming an
obvious choice when the data sources available online were reviewed.

\subsection{Data Source Constraints}

Because ExpertQuest is required to work with real-time data as much
as possible and avoid data extraction, this implies that the data
that will be required for finding experts will most likely be via
one of two methods:
\begin{itemize}
\item Scraping data (text extraction) from websites.
\item Application Programming Interfaces (API) that are exposed by various
organisations on the Internet. 
\end{itemize}
There are others (purchasing formatted data for instance) but they
remain outside the scope of this document.

\subsection{Web Scraping}

This choice was considered as an excellent possibility for extracting
data. For instance, \textit{Google Scholar} contains hundreds of academic
papers and citations. It is an excellent resource for finding experts
in multiple fields, but particularly in the medical and biotechnology
fields. There even exists a Python library for extracting data via
Web scraping from Google Scholar \cite{32}.

However, on further examination, this type of Web scraping breaks
the terms of service of Google Scholar (and multiple other websites
of this type) and the legality of this activity, even for research,
became questionable. Therefore the idea was abandoned. 

Unfortunately Google Scholar does not expose an API, so it was not
an option as a form of data for ExpertQuest. Alongside this decision,
the domains available for ExpertQuest were diminished as there are
not many publicly accessible data sources for academic papers as reliable
as Google Scholar. Other vendors in this area were contacted, but
all they provided were multi-gigabyte data dumps of data that are
several years old. 

\subsection{APIs}

Full lists of what APIs are available on websites like the Programmable
Web \cite{33}. On this website alone there are 13,000+ APIs listed
for everything from 3D to zip codes. However, these APIs break down
into several main categories:
\begin{itemize}
\item APIs provided by commercial companies at a cost.
\item APIs provided by commercial companies available for free but with
limits on usage.

\begin{itemize}
\item A subset of this category are social media APIs.
\end{itemize}
\item APIs provided by non-commercial or academic organisations available
for free.

\begin{itemize}
\item A subset of this category includes most of the Linked Data APIs.
\end{itemize}
\end{itemize}
Unfortunately, the availability of API access is also quite constrained
as many of the commercial organisations who were offering free API
access have removed this access as part of the drive to control their
data as ``big data'' whch is now commercially valuable. \textit{LinkedIn}
is one such business. This service is a professional social network
(\textquotedbl Facebook for work\textquotedbl ) and would be an
excellent source of expertise data. Many professionals effectively
use it as an online curriculum vitae. However, LinkedIn have closed
their API access to all but a couple of partner businesses \cite{34}. 

After reviewing available API data sources and deciding on computer
programming as a good choice for the domain of ExpertQuest, there
were three APIs available that seemed to be useful:
\begin{itemize}
\item \textit{Twitter API}
\item \textit{GitHub API}
\item \textit{DBPedia Linked Data Interface}
\end{itemize}

\subsubsection{Twitter API}

Twitter has an extensive API \cite{35} and as Bozzon et al note,
it is an excellent choice for searching for experts:
\begin{quote}
``Twitter appears {[}to be{]} the most effective social network for
expertise matching, as it very frequently outperforms all other social
networks (either combined or alone)... Twitter appears as well very
effective for matching expertise in domains such as computer engineering,
science, sport, and technology \& games.''\cite{36}
\end{quote}
Twitter offers easy access to data that can be mined for experts and
the API is free, so it is ideal as one of the central data sources
for ExpertQuest.

\subsubsection{GitHub API}

GitHub is the most popular open source code sharing website on the
Internet today and also offers an extensive free API to interact with
the site \cite{37}. As was outlined above, computer programming is
the domain of expertise that ExpertQuest will cover. GitHub makes
an obvious choice as a data source - the site operates almost as a
social media site for developers in many different languages, so the
data stream available in the API is rich and extensive.

\subsubsection{DBPedia Linked Data Interface}

Linked Data is a Semantic Web technology and is an important part
of the hybrid approach as was defined earlier. Linked Data is defined
as a way to use the Web:
\begin{quote}
``to create typed links between data from different sources... ...Linked
Data refers to data published on the Web in such a way that it is
machine-readable, its meaning is explicitly defined, it is linked
to other external data sets, and can in turn be linked to form external
data sets.'' \cite{38}
\end{quote}
Linked Data is important because it allows the meaning of the data
to be explicitly defined so it can be used as reference material by
systems such as ExpertQuest. A system can extract meaningful content
quite easily from the machine readable Linked Data.

Unfortunately after much research, it appears that much of the Linked
Data \cite{39} is not suitable for ExpertQuest as it falls under
the following categories:
\begin{itemize}
\item Data is available on the Web but the server is not reliable enough
to base a system on (for instance, SPARQL end points can disappear
and reappear).
\item The data that is made available is in the format of large data dumps
of RDF triples. This is very useful for most research, but is explicitly
disallowed for ExpertQuest as there will be no complex extracting
or processing of data in this prototype.
\item Much of the data was not in the domain of computer science.
\end{itemize}
The best example of Linked Data available on the Web is also the most
useful for ExpertQuest. \textit{DBPedia} provides a Linked Data interface
to the data that is contained in Wikipedia.

DBPedia provide an API that can be queried in several ways and the
JSON (JavaScript Object Notation) enabled DBpedia Linked Data Interface
\cite{40} was chosen as the best way to get interesting data into
ExpertQuest via the web. The server connection to DBPedia is relatively
reliable and the interface is well documented.

\subsection{Using Functional Programming Language - Clojure }

Several different programming languages were examined for implementing
ExpertQuest, but in the end the decision was made to use the Clojure
programming language \cite{41}. Clojure is a modern functional programming
implementation of Lisp that is written to run on the Java Virtual
Machine (JVM). Clojure offers the following advantages for writing
an expert finding system like ExpertQuest:
\begin{itemize}
\item As it runs on the JVM, it is stable and performs well on multiple
operating systems. All of the extensive Java libraries are also available
to use in Clojure.
\item As Clojure is a Lisp with the ``data as code'' philosophy that this
entails, processing data is extremely easy. Clojure comes with a rich
standard library that makes programming data analysis software very
straightforward. It is also a small language and easy to learn as
a result.
\item Clojure is a modern functional language with data structures that
are immutable by default. This allows the software developer to avoid
many bugs associated with managing state in applications. However,
the language is designed for practical software development and also
comes with an assortment of mutable data types that can be used as
needed.
\item The tooling around Clojure is excellent and it comes with support
for several text editors, integrated development environments (IDEs)
and build/deployment tools.
\end{itemize}
The various Clojure libraries that were used will be discussed in
the next chapter.

\subsection{Twitter Bootstrap}

It was originally envisaged that a complex JavaScript based user interface
would have been desirable for ExpertQuest. There are many programming
languages that compile to JavaScript, including an implementation
of Clojure called ClojureScript, that were considered for this purpose. 

After some research it was concluded that this would complicate the
project beyond the needs of this document. For this prototype a simple
clean and usable Web interface can be produced using the excellent\textit{
Twitter Bootstrap} library \cite{42}. This is a library of HTML and
Javascript components that can be used to build a Web interface quickly
and easily without the major time sink which would be involved in
writing a user interface from scratch.

\subsection{OpenNLP}

For the text analysis portion of the ExpertQuest system, the choice
of libraries was obvious. The \textit{Apache OpenNLP \cite{43}} is
the one of the best machine learning/natural language processing toolkits
available and it is available on the JVM. As Clojure is also hosted
on the JVM, OpenNLP is available to be used in ExpertQuest without
any difficulty. 

\subsection{General Design Approach}

The general design approach for ExpertQuest will be the following:
\begin{itemize}
\item Initially identify candidates via Twitter.
\item Use Linked Data and GitHub data to verify and rank the candidate experts.
\end{itemize}

\subsection{Summary}
\begin{itemize}
\item ExpertQuest will use social media data from the Twitter and GitHub
API for finding experts. 
\item ExpertQuest will also extract content from the Linked Data interface
to DBPedia.
\item ExpertQuest will be written in Clojure, with a Web based interface
using the Twitter Bootstrap library. The OpenNLP libraries will be
used for text processing.
\end{itemize}
The next chapter will go into more detail as to how ExpertQuest is
implemented.

\pagebreak{}

\section{Detailed Design and Implementation}

This chapter contains a detailed description of the design and implementation
of ExpertQuest.

\subsection{Overall System Workflow}

\textbf{Figure 1 }exhibits a diagram for the overall system workflow
for ExpertQuest. The system goes through the following steps to find
experts for a user:
\begin{enumerate}
\item The user chooses a search term in the Web interface. In the prototype,
the user selects a programming language from a drop down list in the
Web interface. The languages presented are the top 50+ languages from
the Tiobe programming language popularity index \cite{44}.
\item The search term is passed to the Twitter Search API.
\item The Twitter Search API returns a number of search results. 

\begin{enumerate}
\item All Twitter usernames in the results are collated and are passed on
to the GitHub API to ascertain if there are matching usernames on
GitHub.
\item All matching GitHub accounts are queried against the GitHub API.
\end{enumerate}
\item All Twitter accounts with matching GitHub accounts are queried using
the Twitter API to obtain a number of most recent tweets.
\item The abstract for the programming language being queried is loaded
from DBPedia via the Linked Data API.
\item The Tweets for each Twitter account are concatenated into a string
and analysed against the DBPedia abstract using a feature hashing
algorithm.
\item All of the results are collated and the expert candidates are ranked
according to several criteria as outlined below.
\end{enumerate}
\begin{figure}
\begin{centering}
\includegraphics[scale=0.5]{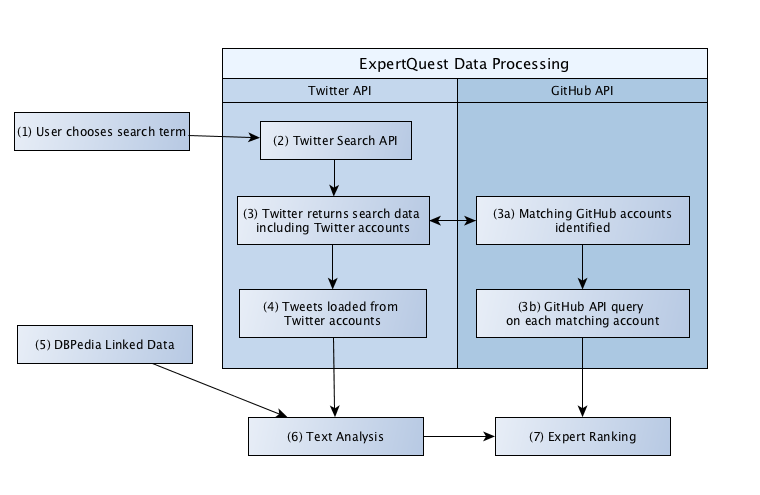}
\par\end{centering}
\caption{\textbf{ExpertQuest Workflow}}
\end{figure}

\subsection{ExpertQuest Namespaces}

Code is abstracted into namespaces in Clojure, which are generally
confined to individual files, similar to Java packaging, each containing
a number of functions. The components mentioned below are each implemented
as individual namespaces in ExpertQuest. 

\subsubsection{Twitter Client}

A simple client for Twitter was written using the recommended Clojure
Twitter API library \cite{45}. There are multiple API calls exposed
by the Twitter API. Only two of these APIs are used in ExpertQuest:
\begin{itemize}
\item The \textit{Twitter Search API} which will return a number of search
results for a specific search term. This is used in step 2 above.
The search term used is the programming language plus the string ``github''.
This is used to disambiguate searches where the programming language
name has also an everyday meaning e.g. Python or Ruby.
\item The \textit{Twitter User Timeline API} which will return a specified
number of most recent tweets for a specified username. This is used
in step 4 above.
\end{itemize}
The Twitter client component wraps all tweet data into a Clojure data
structure and returns it to the search component. The values returned
by the component includes the Twitter user's name, their account name
and the number of followers the user has on Twitter.

\subsubsection{GitHub Client}

A wrapper for the GitHub API was also created using the best available
Clojure GitHub API library \cite{46}. There are again multiple individual
APIs exposed by the GitHub API but there are only two needed by ExpertQuest:
\begin{itemize}
\item A call to the GitHub User API to extract all information pertaining
to one user account. This is used in Step (3a) and (3b) above.
\item A call to the GitHub User Repository API which returns all repositories
belonged to the user. This is used in Step (3b) above and is used
to sum the total bytes of code that are contained in a user's repositories
for a specific language.
\end{itemize}
The GitHub client component wraps all returned GitHub data into a
Clojure data structure and returns it to the search component. The
values returned by the component include the number of followers the
user has on GitHub, and the sum total number of bytes of code in a
specific language that the user has in their Git repositories.

\subsubsection{DBPedia Client}

The Linked Data API of DBPedia provides a REST (Representational State
Transfer) interface that will return JSON data based on a query to
a specific URI. ExpertQuest uses the Clojure HTTP client and JSON
libraries to query this interface for a provided query. The resulting
data is automatically converted to a Clojure data structure and the
abstract (text data) pertaining to the search parameter is extracted.

A call to extract disambiguation data from DBPedia was also created,
but in the end was not necessary for the prototype.

This component is used in step 5.

\subsubsection{Text Analysis Component}

The text analysis component of ExpertQuest is used to measure how
much the expert candidate has been tweeting about the specific programming
language. This component uses the Clojure wrapper library for the
OpenNLP library for this purpose \cite{47}. In order to get a measurement
of how much an expert candidate has been tweeting about a specific
language, the following steps are taken:
\begin{enumerate}
\item The tweets for one specific user that were collected from the Twitter
API in Step 4 are concatenated into one long string and passed to
the text analysis component.
\item The abstract text data from DBPedia (Step 5), about the specific programming
language, is passed to the text analysis component. This text is generally
a few hundred words long and is loaded with descriptive phrases and
terminology pertaining to the programming language. For instance,
here is the Clojure entry:

\begin{quote}
``Clojure (pronounced like \textquotedbl closure\textquotedbl )
is a dialect of the Lisp programming language created by Rich Hickey.
Clojure is a functional general-purpose language, and runs on the
Java Virtual Machine, Common Language Runtime, and JavaScript engines.
Like other Lisps, Clojure treats code as data and has a sophisticated
macro system. Clojure's focus on programming with immutable values
and explicit progression-of-time constructs are intended to facilitate
the development of more robust programs, particularly multithreaded
ones.''
\end{quote}
\item Both pieces of string data are transformed into vectors using the
feature hashing algorithm described below and are compared using cosine
similarity to give a value between 0 and 1, where 1 is more similar.
This value is passed back to be used as part of the expert ranking.
\end{enumerate}

\paragraph{Feature Hashing Algorithm:}

It was originally envisaged that a simple comparison such as the Sorensen
/ Dice coefficient or the Jaro-Winkler distance would have been sufficient
for the text comparison task in ExpertQuest. However these algorithms
are generally used on much shorter strings than those ExpertQuest
is comparing. Further research was carried out to determine a fast
way to compare text documents.

In the field of machine learning and natural language processing,
input data is normally transformed into a \textit{feature vector}
which is an array of integer values where each value represents an
item of a specific type or a measurement of a specific property (a
``feature''). A standard procedure to vectorise a document is to
extract and count the occurrence of each word. This vector is generally
mapped to a document dictionary which is represented as an associative
array where the actual word can be looked up. ExpertQuest needs to
be reasonably fast, and converting text documents into a vector form
for comparison can be slow and use a large amount of RAM \cite{48}. 

Fortunately there is a technique called \textit{feature hashing} available
that will work perfectly for ExpertQuest. Feature hashing is a machine
learning technique (also known as the ``hash trick'') that bypasses
the associative array element of the feature vector. The Wikipedia
article on feature hashing explains this quite succinctly:
\begin{quote}
``{[}feature hashing{]} is a fast and space-efficient way of vectorizing
features, i.e. turning arbitrary features into indices in a vector
or matrix. It works by applying a hash function to the features and
using their hash values as indices directly, rather than looking the
indices up in an associative array.'' \cite{49}
\end{quote}
So rather than store the word in an associative array, we hash the
word and use that resulting value as the index for the word in the
feature vector. This is updated in the vector as appropriate. As Foreman
and Kirshenbaum observed of their implementation of this technique
``SpeedyFX'', the results are compelling and accurate when compared
to the standard method:
\begin{quote}
``We have shown that using SpeedyFx integer hashes in place of actual
words is faster, requires less memory for transmission and use of
multiple classifiers, and has an effect on classification performance
that is practically noise compared to the effect of other common parameters
in model selection.'' \cite{50}
\end{quote}
And again in Weinberger et al, feature hashing is praised for its
benefits:
\begin{quote}
``The benefits of the hashing-trick leads to applications in almost
all areas of machine learning and beyond.'' \cite{51}
\end{quote}
There is no native implementation of the feature hashing technique
in Clojure, so an implementation was written. This implementation
works as follows:
\begin{enumerate}
\item The incoming string parameter is passed to a function which removes
punctuation, filters out everything but nouns and verbs and stems
the remaining words (using the Clojure OpenNLP library).
\item Creates a vector of the remaining words and hashes each one using
a Clojure hashing function.
\item These hash values are stored in an atom (a mutable Clojure data type)
that is updated with each word in turn.
\item The finished vector is passed back as a return.
\end{enumerate}
The actual implementation is displayed in \textbf{Figure 2} above.

\begin{figure}
\begin{lstlisting}[numbers=left,basicstyle={\scriptsize\ttfamily}]
(defn conj-stemmed-nouns-verbs
  "Conj sentences by nouns and verbs, stemming them at the same time"
  [input-text]
  (->> input-text
       (clean-string)
       (nouns-verbs-filter)
       (map #(porter-stem-text (first %1)))
       (string/join " ")))


(defn get-fvec-for-text
  "Transforms a string into a feature hash vector"
  [input-text]
  (clear-fvec-atom)
  (let [stemmed-text (conj-stemmed-nouns-verbs input-text)]
    (let [word-vec (string/split stemmed-text #" ")]
      (doseq [x (range 0 (count word-vec)) :let [word (get word-vec x)]]
        (let [index (mod (digest/crc32 word) fvec-size)]
          (let [old-value (get @fvec-atom  index)]
            (swap! fvec-atom assoc index (inc old-value)))))
      (identity @fvec-atom))))
\end{lstlisting}

\caption{\textbf{Stemming and Feature Hashing Algorithms}}

\end{figure}

The algorithm performs well enough to be usable within a Web interface.
For instance, the text of the novel \textit{Anne of Green Gables}
by Lucy Maud Montgomery, downloaded from Project Gutenburg, contains
about 105,000 words. This was vectorised by the algorithm on a MacBook
Pro with 8gb of RAM in under two minutes. The resulting vector looks
like this: 
\begin{quote}
1442 367 386 50 157 546 193 76 81 379 68 178 103 211 324 153 210 243
38 99 109 66 197 41 130 148 131 262 89 214 54 89 165 210 226 30 153
93 72 51 26 127 123 95 82 949 331 93 109 142 55 41 107 106 37 195
106 72 57 73 59 44 210 95 656 280 46 92 128 425 132 156 53 153 35
67 55 68 141 132 93 519 72 49 62 111 278 49 101 74 103 74 121 124
17 124 95 102 105 113 83 95 116 245 104 165 54 21 224 37 216 122 357
103 297 1336 68 71 215 83 91 34 100 292 42 82 55 86 112 50 118 345
116 589 264 222 23 76 27 58 24 155 89 263 38 114 172 57 139 179 181
311 67 218 60 181 287 33 154 103 71 218 221 900 195 87 117 62 468
149 109 96 116 20 147 244 133 177 88 158 60 47 83 62 95 32 146 329
22 30 60 239 127 31 101 320 193 50 85 67 164 99 170 126 173 215 57
135 353 277 103 131 202 70 680 114 237 526 163 216 168 304 234 104
298 125 298 121 64 82 156 144 59 457 115 89 37 43 54 118 142 75 27
126 329 72 74 97 129 63 149 128 78 99 366 207
\end{quote}
As most of the text documents will be much smaller than this within
ExpertQuest, this level of performance was deemed sufficient for the
prototype.

\paragraph{Cosine Similarity}

The resulting vector for each text document (either a DBPedia abstract
or a string of concatenated tweets) need to be compared for similarity.
The cosine similarity was chosen as it is generally quite fast over
sparse vectors. The actual implementation of the cosine similarity
algorithm was taken from the Clojure Incanter statistical package.
The cosine similarity value is returned back to be used in the ranking
of the expert candidates.

\subsubsection{Search Component}

The search component contains the main search function that is called
when the user submits the form in the Web interface as described below.
This function calls the various components in order as described in
the workflow as set out in \textbf{Figure 1}.

\paragraph{Search Algorithm Constants}

Two constants are defined in the ExpertQuest search component. These
values behave how the search algorithm behaves:
\begin{itemize}
\item \textit{Twitter Search Count}: This is the amount of search items
to load from Twitter. This defaults to 50.
\item \textit{User Timeline Count}: This is the amount of tweets to load
from a user's Twitter timeline for text comparison. This defaults
to 25.
\end{itemize}
This means that for each of the 50 search results that returns a Twitter
account that has a matching GitHub account, 25 tweets will be loaded
from this account for the text analysis. This means that the search
algorithm in ExpertQuest performs at O(n\textasciicircum 2).

\pagebreak{}

\subsubsection{Web Interface}

The ExpertQuest Web interface is developed using the following Clojure
libraries:
\begin{itemize}
\item The Ring library which provides clean abstracted access to a Web server
from within the application \cite{52}.
\item The Compojure library which is used to route URLs internally in the
application \cite{53}.
\item The Hiccup library which is used to output HTML to the client \cite{54}.
\end{itemize}
Twitter Bootstrap is used to provide clean cross-browser typography
and layout (albeit a simple one).

The Web interface is illustrated in the following figures:
\begin{itemize}
\item \textbf{Figure 3,} above, exhibits the search drop down that presents
a list of programming languages. The user can search for programming
experts by selecting a language and clicking the search button. The
data for this drop down list is loaded from a configuration file as
described later in this chapter.
\end{itemize}
\begin{figure}
\begin{centering}
\includegraphics[scale=0.5]{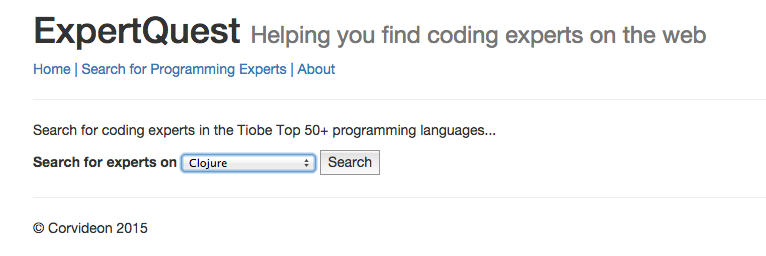}
\par\end{centering}
\caption{\textbf{Search Page}}
\end{figure}

\begin{itemize}
\item \textbf{Figure 4,} overleaf, shows the Twitter Bootstrap modal popup
that the users see when the system is searching. It was decided that
rather than using a more complex asynchronous callback based system,
this user interface was sufficient for the prototype despite its inelegance. 
\end{itemize}
\begin{figure}
\begin{centering}
\includegraphics[scale=0.5]{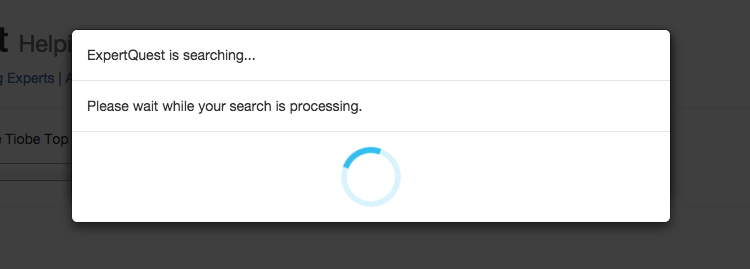}
\par\end{centering}
\caption{\textbf{Searching Modal Popup}}
\end{figure}

\begin{itemize}
\item \textbf{Figure 5,} overleaf, illustrates the search results as displayed
once the search has returned with results. The results are ranked
according to the criteria described later in this chapter. One thing
to note is that the cosine similarity value is listed as the \textit{Twitter
Mentions} column to give the user some indication of how much this
expert candidate has been tweeting about the specific programming
language. This value is displayed as a progress bar, rather than as
a number, for increased visual feedback.
\end{itemize}
\begin{figure}
\begin{centering}
\includegraphics[scale=0.25]{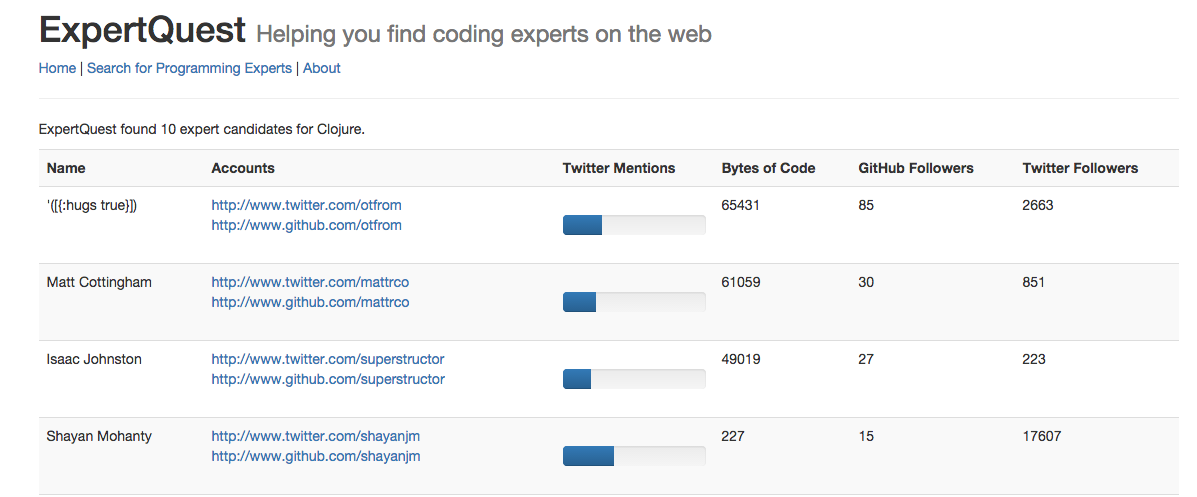}
\par\end{centering}
\caption{\textbf{Search Results}}
\end{figure}

\subsection{Mutable Elements}

As Clojure is a functional programming language, the emphasis is on
minimising mutable state within an application. It is interesting
to note that all the areas where state can be mutated can be listed
quite simply for ExpertQuest as a result (compared to say, an application
written in typical Java, which would have mutable state spread right
through each class).

The mutable state in ExpertQuest consists of the following
\begin{itemize}
\item A Clojure atom that is used when processing the data within the feature
hashing algorithm. 
\item Various Twitter and GitHub security configuration files in the Clojure
EDN format (Extensible Data Notation, basically Clojure data structures
stored as strings in files).
\item An EDN file containing the list of the \textit{Tiobe Top 50+ programming
languages }\cite{55} with the corresponding correct search string
for DBPedia. This was necessary as carrying out a search for ``Java''
in DBPedia returns the island nation rather than the programming language.
The programming language is listed as ``Java (programming language)''.
The author developed code to return disambiguation pages from DBPedia,
but this made the Web interface very slow and the idea was left to
one side for the prototype.
\end{itemize}

\subsection{Expert Ranking}

The expert ranking within ExpertQuest is achieved quite simply thanks
to the rich data handling ability of Clojure. A vector of maps is
passed back to the Web view from the search function and before it
is displayed it is sorted by the following criteria:
\begin{itemize}
\item \textit{Number of Bytes} - The number of bytes of code that the expert
candidate has in their GitHub repositories in the specific programming
language being searched for.
\item \textit{GitHub Followers }- The number of GitHub followers they have
watching their GitHub repositories.
\item \textit{Twitter Mentions} - The cosine similarity of their most recent
tweets when compared with the DBPedia abstract on the specific programming
language being searched for.
\item \textit{Twitter Followers} - The number of Twitter followers they
have in their Twitter account.
\end{itemize}
This code is extracted and displayed in \textbf{Figure 6} above.

\begin{figure}
\begin{lstlisting}[numbers=left,basicstyle={\scriptsize\ttfamily}]
(sort-by 
	(juxt :boc  :github-followers :cosim :twitter-followers))
\end{lstlisting}

\caption{\textbf{Expert Ranking Code Extract}}
\end{figure}

\subsection{Challenges and Compromises}

This section is a discussion of some the challenges that were encountered
during developing ExpertQuest.

\subsubsection{Matching Twitter Accounts with GitHub Accounts}

The system only presents results to the user where the Twitter and
the GitHub name are exactly the same. This is a compromise for the
prototype as it obviously ignores results where the same person has
two different account names on each service.

An early attempt was made to cross reference Twitter and GitHub accounts
using the Yahoo Search API. However, there was no easy way to verify
that any results returned were about the same person - this would
be a project in itself. 

The compromise within ExpertQuest is a reasonable one as each expert
candidate is checked to make sure they were tweeting about a specific
programming language and also that they have code in repositories
in this language. 

\subsubsection{Avoiding Non Programming Homonyms in the Search Results}

Ruby, Python, Java and several other languages are all homonyms. Avoiding
these non programming homonyms in the Twitter Search is a real issue,
so a simple compromise was to also use the word ``github'' in the
search query which helped narrow the results. This is not very elegant,
but it works reasonably well to avoid the homonyms.

\medskip{}

In the next chapter, the ExpertQuest system is evaluated against the
primary and secondary requirements.

\pagebreak{}

\section{Testing and Evaluation}

This chapter outlines the testing and validation carried out on the
results of ExpertQuest.

\subsection{Text Comparison Testing}

The feature hashing / cosine similarity algorithm was tested for accuracy
using some basic tests on the Clojure REPL as illustrated in \textbf{Table
2} overleaf. In each case, string A and B were passed through the
feature hashing algorithm and the resulting vectors were passed through
the cosine similarity algorithm. The higher the score, the more alike
the strings, with the highest possible value being 1.

The scores were predictable and accurate as each test is described
below:
\begin{enumerate}
\item This test shows that the algorithm only compares on nouns and verbs
as designed. These strings are equal as the noun ``dog'' is all
that is being measured.
\item There are three out of four verbs that match in this test with an
expected value of 0.75.
\item The only difference in these strings is the extra verb ``barking'',
so the score is less than one.
\item String A has two nouns and string B has one noun and one verb, hence
the 0.50 score.
\item The difference here is one word out of three (``mouse ran clock''
versus ``elephant ran clock'') so the score is close to two thirds
similar.
\item String A and B are here seen as the same, as word order is irrelevant.
\end{enumerate}
These simple tests show that the text comparison is reasonably accurate.
ExpertQuest only requires reasonable accuracy as the cosine similarity
score is only one factor amongst others for expert ranking (as the
``Twitter Mentions'' measure in the Web interface). 

\subsection{Testing and API Limits}

\subsubsection{Test Runs}

Three different tests were run to measure the performance of ExpertQuest.
The tests took the following format:
\begin{itemize}
\item A data dumper was written that rotated through all 53 programming
languages in the system (see Appendix 1 on page 45) and ran a search
against Twitter and GitHub. The data from these sessions was recorded
and analysed.
\item Each test only differed from the other in the values of the search
component constants as mentioned earlier. Column A in \textbf{Table
3} overleaf is the amount of search items to load from Twitter. Column
B in \textbf{Table 3} overleaf is the amount of tweets to load from
a user's Twitter timeline for text comparison.
\item The upper and lower values for these tests were discovered by trial
and error. The lower value represents the approximate threshold where
some useful values are returned. The upper value represents the approximate
point at which the Twitter and GitHub API limits get hit by the system.
Beyond this point search queries may start failing.
\item The values used in each test run are displayed in \textbf{Table 3}
above. 
\end{itemize}
\begin{table}
\begin{centering}
\begin{tabular}{|c|c|c|c|}
\hline 
Test No. & String A & String B & Cosine Similarity\tabularnewline
\hline 
\hline 
1 & \textquotedbl white dog\textquotedbl{} & \textquotedbl black dog\textquotedbl{} & 1.00\tabularnewline
\hline 
2 & \textquotedbl run jump play hide\textquotedbl{} & \textquotedbl run jump play seek\textquotedbl{} & 0.75\tabularnewline
\hline 
3 & \textquotedbl running dog\textquotedbl{} &  \textquotedbl running and barking dog\textquotedbl{} & 0.82\tabularnewline
\hline 
4 & \textquotedbl runner jump\textquotedbl{} & \textquotedbl running jump\textquotedbl{} & 0.50\tabularnewline
\hline 
5 & \textquotedbl the mouse ran up the clock\textquotedbl{} & \textquotedbl the elephant ran over the clock\textquotedbl{} & 0.67\tabularnewline
\hline 
6 & \textquotedbl the mouse ran up the clock\textquotedbl{} & \textquotedbl the clock ran over the mouse\textquotedbl{} & 1.00\tabularnewline
\hline 
\end{tabular}
\par\end{centering}
\caption{\textbf{Cosine Similarity Results for Example Strings}}
\end{table}

\subsubsection{Defining an Expert for the Test Runs}

ExpertQuest will display all of the data that it returns but will
rank all candidates according to the criteria outlined in section
5.4. Some candidates will have Twitter and GitHub accounts and have
been actively tweeting about the specific programming language in
question. These candidates may be of interest to the user. However,
for the sake of the test runs, only candidates who also have GitHub
repositories containing code in the programming language in question
are considered ``full'' experts. These candidates are counted as
correct returns in the tests as they would be of the most interest
to the user.

\begin{table}
\begin{centering}
\begin{tabular}{|c|c|c|}
\hline 
Test Run No. & A - Twitter Search Count & B - User Timeline Count\tabularnewline
\hline 
\hline 
Test Run 1 & 10 & 5\tabularnewline
\hline 
Test Run 2 & 30 & 15\tabularnewline
\hline 
Test Run 3 & 50 & 25\tabularnewline
\hline 
\end{tabular}
\par\end{centering}
\caption{\textbf{Test Runs}}
\end{table}

\subsection{Test Results}

Please note, more detailed test results are contained in the appendices
on pages 45-49.

\subsubsection{Precision}

The precision of each test was calculated using the number of experts
found divided by the total number of candidates found. As each search
for a programming language is independent, the average precision for
each programming language was worked out across all 53 searches. As
expected, when more data was pulled into the system using the larger
search constants, the precision went up steadily. The average precision
for each test run was as follows (also see \textbf{Figure 7} above):
\begin{enumerate}
\item 0.158265948
\item 0.171069182
\item 0.20215256
\end{enumerate}
\begin{figure}
\begin{centering}
\includegraphics[scale=0.5]{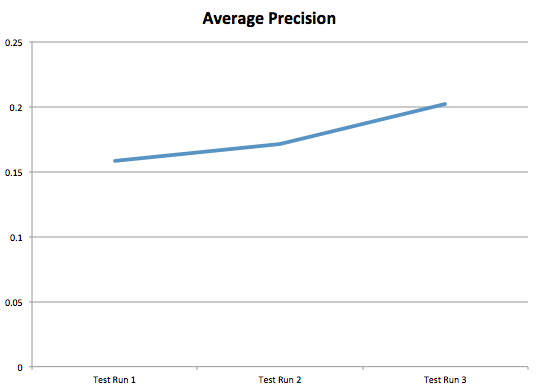}
\par\end{centering}
\caption{\textbf{Average Precision}}
\end{figure}

\subsubsection{Recall}

Each of the test runs passed in a search constant representing the
number of Twitter search results to return (10, 30 or 50). Under ideal
conditions, each of these search results would have contained the
details of an expert. So the maximum number of experts that can ever
be returned is equal to the number of Twitter search results. 

Using this information we can calculate the recall of each test by
dividing the total number of experts found by the Twitter search constant.
As illustrated above, the results of each test were calculated and
then averaged across all 53 programming languages. The average recall
for each test run was as follows (also see \textbf{Figure 8,} overleaf):
\begin{enumerate}
\item 0.052830189 
\item 0.020754717 
\item 0.050943396
\end{enumerate}
The recall dips in the second test run even though the search constants
were increased, the number of experts returned was not more much than
in first test. The third test was much better, though surprisingly
at a level similar to the first test. A lot more candidates were returned
in the third test, but the ratio of experts to non-experts does not
increase. This suggests that even with more data being added to the
algorithm, the percentage of relevant returns (experts in this case)
does not increase significantly.

\begin{figure}
\begin{centering}
\includegraphics[scale=0.5]{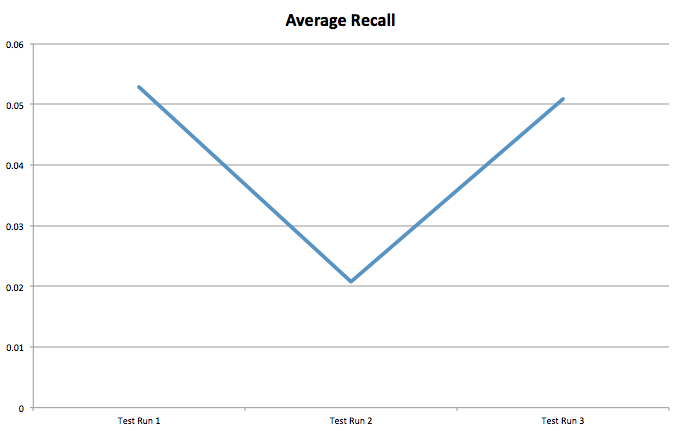}
\par\end{centering}
\caption{\textbf{Average Recall}}
\end{figure}

\subsubsection{Cosine Similarity}

ExpertQuest measures the cosine similarity of each candidates tweets
and the DBPedia abstract on the programming language, as laid out
in section 5.2.4 on page 28. This is used to represent how much each
candidate was tweeting about the specific programming language. As
expected, the average cosine similarity goes up across each test run
as the amount of tweet data is increased. This can be seen in \textbf{Figure
9,} overleaf.

\begin{figure}
\begin{centering}
\includegraphics[scale=0.5]{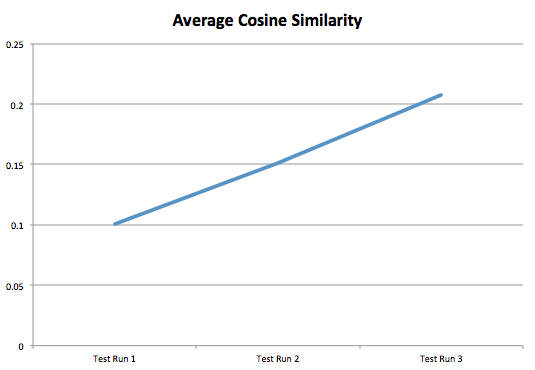}
\par\end{centering}
\textbf{\caption{\textbf{Average Cosine Similarity}}
}
\end{figure}

\subsubsection{Comparison of Different Programming Languages}

The results for Test Run 3 are displayed in \textbf{Table 4}. These
results represent the ExpertQuest at its most useful. Some results
that are revealing:
\begin{itemize}
\item ExpertQuest was only able to find experts in 21 out of the 53 programming
languages.
\item ExpertQuest was most successful at finding experts in programming
languages whose development is actually hosted on GitHub. The precision
here was much higher and the results much more relevant. The top four
languages where experts were found are JVM (Jave Virtual Machine)
languages.
\end{itemize}
\begin{table}
\begin{centering}
\begin{tabular}{|c|c|c|}
\hline 
 & Total Candidates & Experts\tabularnewline
\hline 
\hline 
Scala & 18 & 14\tabularnewline
\hline 
Clojure & 22 & 14\tabularnewline
\hline 
Groovy & 24 & 14\tabularnewline
\hline 
Java & 17 & 12\tabularnewline
\hline 
Erlang & 18 & 11\tabularnewline
\hline 
Python & 16 & 10\tabularnewline
\hline 
Ruby & 15 & 9\tabularnewline
\hline 
Haskell & 13 & 8\tabularnewline
\hline 
Perl & 12 & 7\tabularnewline
\hline 
Dart & 16 & 7\tabularnewline
\hline 
Lua & 18 & 7\tabularnewline
\hline 
C & 11 & 4\tabularnewline
\hline 
Go & 28 & 4\tabularnewline
\hline 
R & 9 & 3\tabularnewline
\hline 
C++ & 11 & 3\tabularnewline
\hline 
C\# & 8 & 2\tabularnewline
\hline 
D & 23 & 2\tabularnewline
\hline 
AWK & 1 & 1\tabularnewline
\hline 
MATLAB & 1 & 1\tabularnewline
\hline 
Prolog & 2 & 1\tabularnewline
\hline 
Smalltalk & 5 & 1\tabularnewline
\hline 
\end{tabular}
\par\end{centering}
\caption{\textbf{Programming Language Results from Test Run 3}}

\end{table}

\subsection{Achievement of Requirements}

ExpertQuest can now be evaluated as to whether it meets the primary
and secondary requirements as outlined in Chapter 3. 

\subsubsection{Primary Requirements}

These requirements are the key requirements that Maybury \cite{56}
outlines for an expert finding system.
\begin{itemize}
\item \textbf{The system should be able to identify experts from a field
of candidates: }ExpertQuest meets this requirement. For the test runs,
experts were defined as candidates with GitHub repositories in the
desired programming language (see section 6.2.2 page 37) and ExpertQuest
will rank these individuals higher than those without such repositories.
\item \textbf{The system should be able to classify the level of expertise
of each candidate: }ExpertQuest meets this requirement. The display
of candidates offers several measures that can be used to classify
the candidates - the number of Twitter and GitHub followers, the amount
of code in their repositories and the amount they have been tweeting
about the desired programming language.
\item \textbf{The system should be able to validate the expertise of each
candidate: }ExpertQuest meets this requirement. GitHub followers are
a sign that a candidate's repositories are useful and is a reasonable
way of assessing the candidate's credibility. A link to the candidates
GitHub account is also supplied in the interface, so the user can
go and take a look at the code for themselves.
\item \textbf{The system should be able to rank candidates on multiple dimensions:}
ExpertQuest meets this requirement. Expert candidates are ranked simply
in descending order according to the criteria outlined in section
5.4 (on page 34).
\end{itemize}

\subsubsection{Secondary Requirements}
\begin{itemize}
\item \textbf{The system should use the hybrid approach:} As outlined in
section 3.1 (on page 19), the hybrid approach uses all three approaches
in one system - content analysis, social graph analysis and Semantic
Web technology.

\begin{itemize}
\item Content analysis is done on the Twitter stream and DBPedia abstracts
via feature hashing.
\item Social graph analysis is provided via information on followers in
the Twitter and GitHub APIs, albeit in a very trivial form.
\item Semantic Web technology is used in the form of the Linked Data interface
with DBPedia.
\end{itemize}
\item \textbf{The system should measure in some meaningful way all three
properties of expertise: }ExpertQuest meets this requirement as follows:

\begin{itemize}
\item A candidate can have tweets, repositories and GitHub followers, demonstrating
knowledge and credibility within the programming language in question.
\item A candidate can have followers both on Twitter and GitHub, demonstrating
that their social peers are engaged with their expertise.
\end{itemize}
\item \textbf{The system should be based on real-time data available on
the Internet and should also minimise any need for expensive extraction
and transformation of data:} ExpertQuest only uses live data from
Web based APIs.
\item \textbf{The system should also allow a way to contact the expert if
possible: }ExpertQuest provides links to the candidate's Twitter and
GitHub profiles.
\item \textbf{The system should be a usable prototype: }The finished system
returns experts in approximately 40\% of the languages used in the
test runs. This is a weak but positive result.
\item \textbf{If possible, the system should be user friendly and reasonably
fast and should have a Web interface: }The system is not fast but
it is Web based and easy to use.
\end{itemize}

\subsection{Summary}
\begin{itemize}
\item As more Twitter data was added to ExpertQuest the average precision
and average cosine similarity went up. Recall was not increased by
adding more data.
\item ExpertQuest returns useful and relevant results, but testing suggests
that this happens only in a specific subset of languages which are
hosted on GitHub.
\item ExpertQuest meets the requirements outlined in Chapter 3 with some
caveats.
\end{itemize}
\newpage{}

\section{Conclusions and Future Work}

In Chapter 2, expert finding was defined as efficiently identifying
the right individual (or group) from a field of candidates that has
the expertise to provide desired information or complete a desired
task. As we have seen, ExpertQuest did provide an efficient way to
find experts in specific programming languages across Twitter and
GitHub. ExpertQuest will now be analysed as the prototype for a possible
commercial system.

\subsection{SWOT Analysis}

\subsubsection{Strengths}
\begin{itemize}
\item ExpertQuest met the requirements for an expert system as set out in
Chapter 3.
\item ExpertQuest demonstrates that one can build a reasonably useful system
that returns experts in a specific field using only data available
via API. For programming languages that have large communities on
GitHub, the experts identified are often the programmers who play
major roles in their respective language ecosystems as library authors
and the like.
\item The Clojure language and tooling infrastructure is an excellent choice
for this kind of Web enabled data extraction project. The Lisp based
``data as code'' philosophy and the immutable data structures makes
the code succinct and concise. Furthermore, the fact that Clojure
has built-in concurrency means that ExpertQuest could scale if needed.
The available of Java based libraries such as OpenNLP is also a major
advantage.
\end{itemize}

\subsubsection{Weaknesses}
\begin{itemize}
\item ExpertQuest is pushing the limits for how slow a Web interface can
be and still remain useful. Queries may take several minutes. A different
design may be warranted if commercial development was taken further
- batching and processing results in a database perhaps. This would
break one of the requirements for the prototype (no data extraction).
\item Over 60\% of the programming languages returned no useful results
(though this is a factor of the data sources chosen, some programming
languages are just not discussed on Twitter). The precision and recall
results are poor for many of the languages that do return results
also.
\item The first large compromise in the design as outlined in section 5.5
(matching Twitter and GitHub accounts, see pages 34-35) is a major
weakness. In reality a commercial software product with this flaw
would not be useful as it would miss large chunks of data. Even the
solution used here has a small chance of associating two different
people in the same profile. This would need to be remedied before
a commercial product could be released.
\end{itemize}

\subsubsection{Opportunities}
\begin{itemize}
\item The general design approach defined in section 4.8 (page 25) could
be reused for other areas of expertise i.e. Twitter can be used as
a way of identifying expert candidates that could then be filtered
and ranked via information from other sources.
\end{itemize}

\subsubsection{Threats}
\begin{itemize}
\item ExpertQuest is built on the APIs of a company that is notorious for
cutting off access to developers. Twitter has killed multiple startup
businesses in the last few years in this way. 
\item API limits are also a factor and if the application became popular
this could become a real bottleneck.
\item Privacy and data protection could also become major issues if a commercial
version of ExpertQuest was released.
\end{itemize}

\subsection{Future Work}

There are several areas of possible future work that arise out of
this research:
\begin{itemize}
\item Researching how one authenticates the accounts of one individual across
multiple social networks. Perhaps a technology such as FOAF could
be used to solve this problem. 
\item The general approach of augmenting Twitter and other social network
data with other sources of data could also be explored in other areas
such as biotechnology. If a reliable source of up-to-date scientific
research was made available, the hybrid approach used by ExpertQuest
could then be used to create a system for expert finding using this
data source to rank expertise.
\item ExpertQuest could be further developed as full application using a
data store of some kind. This would mean that the data would be slightly
out of date but the queries would run a lot faster. There would also
be opportunities for doing further analysis across various programming
languages communities, for instance, seeing how experts and their
followers are related to one another across the social graph. There
are multiple projects of this type that could be carried out using
a few weeks worth of ExpertQuest queries.
\end{itemize}
\pagebreak{}

\section*{Appendix 1: List of Tiobe Top 50+ Programming Languages}
\begin{itemize}
\item ABAP
\item ActionScript
\item Ada
\item Assembly language
\item AWK
\item Bash
\item C
\item C\#
\item C++
\item Clojure
\item COBOL
\item CoffeeScript
\item D
\item Dart
\item Eiffel
\item Erlang
\item F\#
\item Forth
\item Fortran
\item FoxPro
\item Go
\item Groovy
\item Haskell
\item Inform
\item Java
\item JavaScript
\item LabVIEW
\item Lisp
\item Logo
\item Lua
\item MATLAB
\item Max
\item ML
\item Objective-C
\item OpenEdge ABL
\item Pascal
\item Perl
\item PHP
\item PL/I
\item PL/SQL
\item PostScript
\item Prolog
\item Python
\item R
\item RPG
\item Ruby
\item Scala
\item Scheme
\item Scratch
\item Smalltalk
\item T-SQL
\item VB
\item VB .NET
\end{itemize}
\newpage{}

\section*{Appendix 2: Test Run 1 Results }

\begin{tabular}{|c|c|c|c|c|}
\hline 
Programming Language & Total Candidates Found & Total Experts{*} Found & Precision & Recall\tabularnewline
\hline 
\hline 
AWK & 1 & 1 & 1.000 & 0.100\tabularnewline
\hline 
Dart & 2 & 2 & 1.000 & 0.200\tabularnewline
\hline 
Perl & 2 & 2 & 1.000 & 0.200\tabularnewline
\hline 
Java & 4 & 3 & 0.750 & 0.300\tabularnewline
\hline 
MATLAB & 3 & 2 & 0.667 & 0.200\tabularnewline
\hline 
Python & 3 & 2 & 0.667 & 0.200\tabularnewline
\hline 
Erlang & 5 & 3 & 0.600 & 0.300\tabularnewline
\hline 
Scheme & 5 & 3 & 0.600 & 0.300\tabularnewline
\hline 
Go & 4 & 2 & 0.500 & 0.200\tabularnewline
\hline 
Prolog & 2 & 1 & 0.500 & 0.100\tabularnewline
\hline 
Clojure & 7 & 2 & 0.286 & 0.200\tabularnewline
\hline 
Haskell & 7 & 2 & 0.286 & 0.200\tabularnewline
\hline 
C++ & 5 & 1 & 0.200 & 0.100\tabularnewline
\hline 
Groovy & 6 & 1 & 0.167 & 0.100\tabularnewline
\hline 
Scala & 6 & 1 & 0.167 & 0.100\tabularnewline
\hline 
 &  &  &  & \tabularnewline
\hline 
\end{tabular}

{*}See section 6.2.2 for expert definition

\newpage{}

\section*{Appendix 3: Test Run 2 Results }

\begin{tabular}{|c|c|c|c|c|}
\hline 
Programming Language & Total Candidates Found & Total Experts{*} Found & Precision & Recall\tabularnewline
\hline 
\hline 
AWK & 1 & 1 & 1.000 & 0.033\tabularnewline
\hline 
Dart & 1 & 1 & 1.000 & 0.033\tabularnewline
\hline 
Perl & 2 & 2 & 1.000 & 0.067\tabularnewline
\hline 
R & 3 & 3 & 1.000 & 0.100\tabularnewline
\hline 
Ruby & 5 & 4 & 0.800 & 0.133\tabularnewline
\hline 
Erlang & 5 & 3 & 0.600 & 0.100\tabularnewline
\hline 
Java & 5 & 3 & 0.600 & 0.100\tabularnewline
\hline 
Scheme & 5 & 3 & 0.600 & 0.100\tabularnewline
\hline 
Groovy & 6 & 3 & 0.500 & 0.100\tabularnewline
\hline 
Clojure & 5 & 2 & 0.400 & 0.067\tabularnewline
\hline 
MATLAB & 5 & 2 & 0.400 & 0.067\tabularnewline
\hline 
Go & 6 & 2 & 0.333 & 0.067\tabularnewline
\hline 
Haskell & 6 & 2 & 0.333 & 0.067\tabularnewline
\hline 
Python & 3 & 1 & 0.333 & 0.033\tabularnewline
\hline 
Scala & 6 & 1 & 0.167 & 0.033\tabularnewline
\hline 
 &  &  &  & \tabularnewline
\hline 
\end{tabular}

{*}See section 6.2.2 for expert definition

\newpage{}

\section*{Appendix 4: Test Run 3 Results }

\begin{tabular}{|c|c|c|c|c|}
\hline 
Programming Language & Total Candidates Found & Total Experts{*} Found & Precision & Recall\tabularnewline
\hline 
\hline 
AWK & 1 & 1 & 1.000 & 0.020\tabularnewline
\hline 
MATLAB & 1 & 1 & 1.000 & 0.020\tabularnewline
\hline 
Scala & 18 & 14 & 0.778 & 0.280\tabularnewline
\hline 
Java & 17 & 12 & 0.706 & 0.240\tabularnewline
\hline 
Clojure & 22 & 14 & 0.636 & 0.280\tabularnewline
\hline 
Python & 16 & 10 & 0.625 & 0.200\tabularnewline
\hline 
Haskell & 13 & 8 & 0.615 & 0.160\tabularnewline
\hline 
Erlang & 18 & 11 & 0.611 & 0.220\tabularnewline
\hline 
Ruby & 15 & 9 & 0.600 & 0.180\tabularnewline
\hline 
Groovy & 24 & 14 & 0.583 & 0.280\tabularnewline
\hline 
Perl & 12 & 7 & 0.583 & 0.140\tabularnewline
\hline 
Prolog & 2 & 1 & 0.500 & 0.020\tabularnewline
\hline 
Dart & 16 & 7 & 0.438 & 0.140\tabularnewline
\hline 
Lua & 18 & 7 & 0.389 & 0.140\tabularnewline
\hline 
C & 11 & 4 & 0.364 & 0.080\tabularnewline
\hline 
R & 9 & 3 & 0.333 & 0.060\tabularnewline
\hline 
C++ & 11 & 3 & 0.273 & 0.060\tabularnewline
\hline 
C\# & 8 & 2 & 0.250 & 0.040\tabularnewline
\hline 
Smalltalk & 5 & 1 & 0.200 & 0.020\tabularnewline
\hline 
Go & 28 & 4 & 0.143 & 0.080\tabularnewline
\hline 
D & 23 & 2 & 0.087 & 0.040\tabularnewline
\hline 
 &  &  &  & \tabularnewline
\hline 
\end{tabular}

{*}See section 6.2.2 for expert definition

\pagebreak{}

\end{document}